\documentstyle[12pt,epsf]{article}
\def\beq{\begin{equation}}
\def\eeq{\end{equation}}

\def\beqn{\begin{eqnarray}}
\def\eeqn{\end{eqnarray}}
\relax
\begin{document}
\begin{titlepage}
\def\ba{\begin{array}{c}}
\def\ea{\end{array}}
\def\thefootnote{\fnsymbol{footnote}}
\vfill
\hskip 4in BNL-61907

\hskip 4in ISU-HET-95-4

\hskip 4in June, 1995
\vskip 1.2in
\begin{center}
{\large \bf LIMITS ON NON STANDARD TOP QUARK COUPLINGS FROM
ELECTROWEAK MEASUREMENTS}\\

\vspace{1 in}
{\bf S.~Dawson$^{(a)}$}\footnote{This manuscript has been authored
under contract number DE-AC02-76CH00016 with the U.S. Department
of Energy.  Accordingly, the
U.S. Government retains a non-exclusive, royalty-free license to
publish or reproduce the published form of this contribution, or
allow others to do so, for U.S. Government purposes.}
{\bf  and G.~Valencia$^{(b)}$}\\
{\it  $^{(a)}$ Physics Department,
               Brookhaven National Laboratory,  Upton, NY 11973}\\
{\it  $^{(b)}$ Department of Physics,
               Iowa State University,
               Ames IA 50011}\\
\vspace{1 in}
\end{center}
\begin{abstract}
We calculate the typical size of
loop corrections to electroweak observables arising from
non-standard $Z {\overline t } t$ and $W t b$ vertices.  We use an
effective Lagrangian formalism based on the electroweak
gauge group $SU(2)_L\times U(1)_Y \rightarrow U(1)_{EM}$.  Limits
on the non-standard model top quark couplings from electroweak observables
are presented and compared with previously obtained limits.

\end{abstract}
\end{titlepage}
\clearpage

\section{Introduction}

The large mass of the recently discovered top quark
suggests that the top quark is fundamentally different
from the five lighter quarks.  A number of models
have been proposed in which the interactions of the top quark
are  responsible for the $SU(2)_L\times U(1)_Y$ electroweak
symmetry breaking \cite{ttbar,hill}.
In this note, we make a model-independent study of
the limits on the couplings of the third generation quarks to gauge bosons
 which can be inferred from  the LEP and SLC data.

Precision electroweak measurements  at LEP  and SLC
offer a window into the $Z
{\overline t} t$ and $W tb$ couplings since these
couplings appear in
loop corrections to the $Z$ decay widths.
  Rare $B$ and $K$ decays such as $b\rightarrow s
\gamma$ can also probe the $W tb$ coupling \cite{rare},
while in the future, the $W tb$ coupling
can perhaps be measured to an accuracy of around $30\%$ at the
Tevatron through single top production \cite{yuan} .

As has been emphasized recently \cite{shif},
 the determination of $\alpha_s(M_Z)$ at LEP is somewhat
higher than that obtained from low energy measurements such as deep
inelastic scattering.
The existence of new non-standard
model physics will affect the extraction of $\alpha_s$ at the $Z$
pole and in many cases will lower it to be in agreement with
the low energy
data \cite{as,lang}.


We consider a picture in which  the new physics responsible for the
electroweak symmetry breaking is at some high scale, $\Lambda >
4 \pi v \sim  3~TeV$.
In this case, the physics at low energy can be written in terms of an
effective Lagrangian
describing the interactions of
the $SU(2)_L\times U(1)_Y$ gauge fields with the
Goldstone bosons which become the longitudinal components of the
$W$ and $Z$ gauge bosons.
  It is straightforward to incorporate
fermions into this picture \cite{yuan,pz,dv}.
We will not concern ourselves with
the source of the new physics--it could be supersymmetry, technicolor,
top color, or something else entirely.
The only important point is that
 the new physics occurs at a high scale, so
that an expansion in powers of $M_Z^2/\Lambda^2$ is
appropriate.

\section{Effective Lagrangian}

In this note, we assume that whatever is responsible for generating
the non-standard model  top quark
 couplings occurs at a high scale (perhaps
$\Lambda \sim 3~TeV$) and so the use of a non-linear effective Lagrangian
is appropriate.
This Lagrangian can be used to describe the electroweak sector of
the theory at low energy.
We will assume that only the top quark couplings
are non-standard; the case of non-standard $b$- quark couplings
to the $Z$ has been examined by many authors \cite{zbb}.
If we consider the gauge group $SU(2)_L\times U(1)_Y$
broken to $
U(1)_{EM}$, then there are $3$ Goldstone bosons, $\omega_i$, which
eventually become the longitudinal components of the $W^\pm$ and $Z$
gauge bosons.  The minimal
Lagrangian which describes the interactions of the $SU(2)_L\times U(1)_Y$
gauge bosons with the Goldstone bosons  has been given in \cite{longo}.
  This non-renormalizable
Lagrangian is interpreted as an effective field theory, valid below
some scale $\Lambda$
and  yields the gauge boson self
interactions that we use in this calculation \cite{dv}.

In a similar way one can write an effective Lagrangian that
describes the interactions of fermions to the electroweak gauge
bosons. The terms that are needed for our calculation
have been described in the literature \cite{pz}
and in unitary gauge they
are:
\beqn
{\cal L}_{\rm eff} & = & -{g\over 2 \cos \theta_W}
\biggl[ (L_t+\delta L_t) {\overline t}_L \gamma^\mu t_L
+(R_t+\delta R_t) {\overline t}_R \gamma^\mu t_R
\biggr] Z_\mu \nonumber \\
&&-\biggl\{ {g\over \sqrt{2}}
\biggl[(1+\delta \kappa_L){\overline t}_L \gamma^\mu b_L+
(\delta \kappa_R) {\overline t}_R \gamma^\mu b_R
\biggr] W_\mu^+ + h.c.\biggr\}
\label{ferml}
\eeqn
where we use the standard notation
$t_{L,R}=(1\mp \gamma_5)/2 t$,
$L_t  =  1-4/3 s_Z^2$,
$R_t  =  -4/ 3 s_Z^2$.
We have assumed that the new interactions are CP conserving
and so there are 4 new real parameters to be examined:
$\delta L_t$, $\delta R_t$, $\delta \kappa_L$, and $\delta \kappa_R$.
We will use Eq.~\ref{ferml} at one-loop, but we will not consider
other anomalous couplings that may act as counterterms to this
one-loop calculation. Therefore, our results will not constitute
strict bounds on the parameters of Eq.~\ref{ferml}, but instead
they will depend on the naturalness assumption that contributions
from different couplings do not cancel each other. For example,
we do not discuss possible flavor changing neutral current couplings.

\section{Limits from Electroweak data}

Possible effects of new physics at LEP
and SLC from non-standard $t$-quark couplings
can be parameterized in terms of the $4$ parameters
$S$, $T$, $U$, and $\delta_{b {\overline b}}$ \cite{stu}.
The parameters $S$, $T$, and $U$ describe the effects of the
new physics on the gauge boson two point functions, while
$\delta_{b {\overline b}}$ contains the non-oblique corrections
to  the $Z\rightarrow b {\overline b}$ decay rate.  The non-standard top
quark couplings do not generate non-oblique corrections to decay rates other
than $Z\rightarrow b\overline{b}$.\footnote{There are of course
non-oblique corections to decay rates such as $Z\rightarrow
d \overline{d}$ coming from the $Zt{\overline t}$ vertex, but these
are suppressed by small KM mixing angles, $\mid V_{td}\mid^2$,
and so we neglect them.}

We perform a one-loop calculation of these effects
and retain terms linear and quadratic in the small parameters
($\delta \kappa_L,\delta \kappa_R, \delta L_t$, and
$\delta R_t$). Of course, our formalism is based on the assumption that
the new couplings are small corrections to the minimal standard model
couplings and thus the linear terms dominate the quadratic terms
whenever our formalism is valid. We keep the quadratic terms only because
there is one coupling that does not contribute linearly to the processes
in question. Since the bounds we obtain are based on naturalness
assumptions, among them that there are no cancellations between
different contributions to the physical observables, they can be
applied to this quadratic term as well.

Our one loop calculation including the new couplings is divergent.
In the language of effective field theories these divergences
would be absorbed by higher dimension counterterms. We will make
use of the divergent terms to place bounds on the couplings by
examining the leading non-analytic contribution to the amplitudes
as discussed in Ref.~\cite{dv}. To this effect we first regularize the
integrals in $n=4-2\epsilon$ space-time dimensions.  The
coefficient of $1/\epsilon$ is also the coefficient of the term
$\log\mu^2$ and thus gives the leading non-analytic term.
The use of these terms as estimates for the size of counterterms
has been emphasized in Ref.~\cite{georgi}.

The caveats discussed in Ref.~\cite{dv} apply here as well: our
bounds are valid only as order of magnitude estimates. They are
also based on the assumption that terms induced by different
couplings do not cancel against each other.
In other words, the LEP observables do not directly measure the coefficients
in Eq. \ref{ferml} and it is only from naturalness arguments that
we can place limits on the anomalous top quark couplings.  For this
reason our bounds are not a substitute for direct measurements
in future high energy machines.

In order to obtain limits we use the fit of Ref.~\cite{lang} to
possible ``new physics'' contributions to
$S$,$T$, $U$, and $\delta_{b {\overline b}}$.  This fit assumes
$M_t = 175 \pm 16 ~GeV$ and gives:
\beqn
S_{\rm NEW}^{exp}& =&-.21\pm .24^{-.08}_{+.17}\nonumber \\
T_{\rm  NEW}^{exp}&=&-.09\pm .32^{+.16}_{-.11}\nonumber \\
U_{\rm  NEW}^{exp}& =&-.53\pm .61\nonumber \\
\delta_{b {\overline b},{\rm NEW}}^{exp}&=&.022\pm .011
\quad .\nonumber \\
\label{langnum}
\eeqn
The standard model top quark  and Higgs boson
contributions are ${\bf {\it not}}$
included in the ``new physics'' contributions.  The
central values in Eq. \ref{langnum} are for $M_H=300~GeV$, the upper
second errors are the differences when
 $M_H=1~TeV$, while the lower second errors
take $M_H=60~GeV$.

Refs.~\cite{yuan,pz,zhang} computed the contributions of
${\cal O}({M_t^2/M_W^2})$ to $T$ and $\delta_{b {\overline b}}$. Here we
extend that calculation to include all the leading non-analytic terms,
even those not enhanced by $M_t^2/M_W^2$.
We find
\beqn
S_{\rm NEW}&=& {1\over 3 \pi} \biggl[
 2\delta R_t- \delta L_t-{3\over 2}(
\delta L_t^2 +\delta R_t^2)\biggr]
\log ({\mu^2\over M_Z^2})\nonumber \\
T_{\rm NEW} &=& {3\over 4 \pi s_Z^2}
\biggl({M_t\over M_W}\biggr)^2 \biggl[
\delta \kappa_L +{\delta \kappa_L^2
+\delta \kappa_R^2\over 2} +\delta R_t -\delta L_t
\nonumber \\  &&
-{(\delta L_t-\delta R_t)^2\over 2}\biggr]
\log ({\mu^2\over M_Z^2}) \nonumber \\
U_{\rm NEW}&=&{1\over \pi} \biggl[ -2\delta \kappa_L
-(\delta\kappa_L^2+\delta \kappa_R^2) +
           \delta L_t
+{\delta L_t^2+\delta R_t^2\over 2}\biggr]
\log ({\mu^2\over M_Z^2})\nonumber \\
\delta_{b {\overline b}, {\rm NEW}}&=&
{\alpha\over 9 \pi} {1\over L_B^2 +R_B^2}\biggl\{
{3 (4 s_Z^4-18 s_Z^2+9)\over 4 s_Z^2 \cos 2 \theta_W}
{M_t^2\over M_W^2} \biggl[-{(\delta R_t-\delta L_t)^2\over 2}
+{\delta \kappa_L^2\over 2} +\delta \kappa_L
\biggr]\nonumber \\  &&
+{3\over 4 s_Z^2 \cos 2 \theta_W }
{M_t^2\over M_W^2}\biggl[
\delta L_t(4 s_Z^4+2 s_Z^2-3)
-{\delta\kappa_R^2\over 2}(28 s_Z^4+2 s_Z^2-9)\nonumber \\&&
+{\delta R_t\over 2}(4 s_Z^4-28 s_Z^2 +15)\biggr]
+{4 s_Z^2-3\over 3 \cos 2 \theta_W}\biggl( 2 \delta R_t-\delta L_t
-{3\over 2}(\delta L_t^2+\delta R_t^2)\biggr)
\nonumber \\ &&
+{(16 s_Z^2-17)\over 4 c_Z^2 s_Z^2}
\biggl[\biggl({\delta \kappa_L^2\over 2}+\delta \kappa_L
\biggr)(2 s_Z^2-3) +\delta \kappa_R^2 \biggr]
\biggr\}
\log ({\mu^2\over M_Z^2}),
\eeqn
where $L_B= -1+2/ 3s_Z^2$  and $R_B= 2/ 3s_Z^2$.
We have not included contributions that are independent of
$\delta \kappa_L$,
$\delta \kappa_R$,
$\delta L_t$, or $\delta R_t$. These were the subject of Ref.~\cite{dv}.

As discussed in Ref.~\cite{dv}, we choose as input parameters for our
renormalization scheme $G_F$, $M_Z$ and $\alpha(M_Z)=1/128.8$. In this
scheme $M_W$ and $s_Z^2$ are derived quantities. Numerically:
\beqn
\delta_{b {\overline b}}&=&{\alpha\over \pi}
\biggl\{
{M_t^2\over M_W^2}\biggl[
-2.29\biggl((\delta R_t-\delta L_t)^2
-\delta \kappa_L^2-2\delta \kappa_L\biggr)\nonumber \\
&&+3.19\delta \kappa_R^2-2.11\delta L_t
+3.96\delta R_t\biggr]
+.29(\delta R_t^2+\delta L_t^2)+.19(\delta L_t-2\delta R_t)
\nonumber \\  &&
-.65\delta\kappa_R^2+3.57(\delta \kappa_L^2+2\delta\kappa_L)
\biggr\}
\log({\mu^2\over M_Z^2})
\quad .
\label{bbnum}
\eeqn

The anomalous couplings are assumed to be small, so we will
retain only the linear couplings with the exception of
$\delta \kappa_R$ as mentioned before. We follow the philosophy
of Ref.~\cite{georgi} in our use of the leading non-analytic terms.
Accordingly, we must choose a renormalization scale below the
scale of new physics $\Lambda \approx 3~ TeV$ in such a way that
the logarithms are of order one. With the prejudice that these
new couplings are somehow related to the top-quark mass, we
choose for our estimates $\mu = 2 M_t$. Using
$M_t=175\pm 16~GeV$ we find the $90\%$ confidence level limits:
\begin{itemize}

\item From $U^{exp}_{\rm NEW}$:\quad
$-2 < -2\delta \kappa_L+\delta L_t < .6$
\item From $S^{exp}_{\rm NEW}$:\quad
$-2 < 2 \delta R_t -\delta L_t < 1$
\item From $T^{exp}_{\rm NEW}$ :\quad
$-.06 < \delta \kappa_L+\delta R_t -\delta L_t < .05$
\item From $\delta_{b\overline{b},{\rm NEW}}^{exp}$:\quad
$.02 < \delta \kappa_L-.3\delta L_t +.6 \delta R_t < .2$

\end{itemize}

\epsfysize=4.in
\begin{figure}
{\centerline{\epsffile{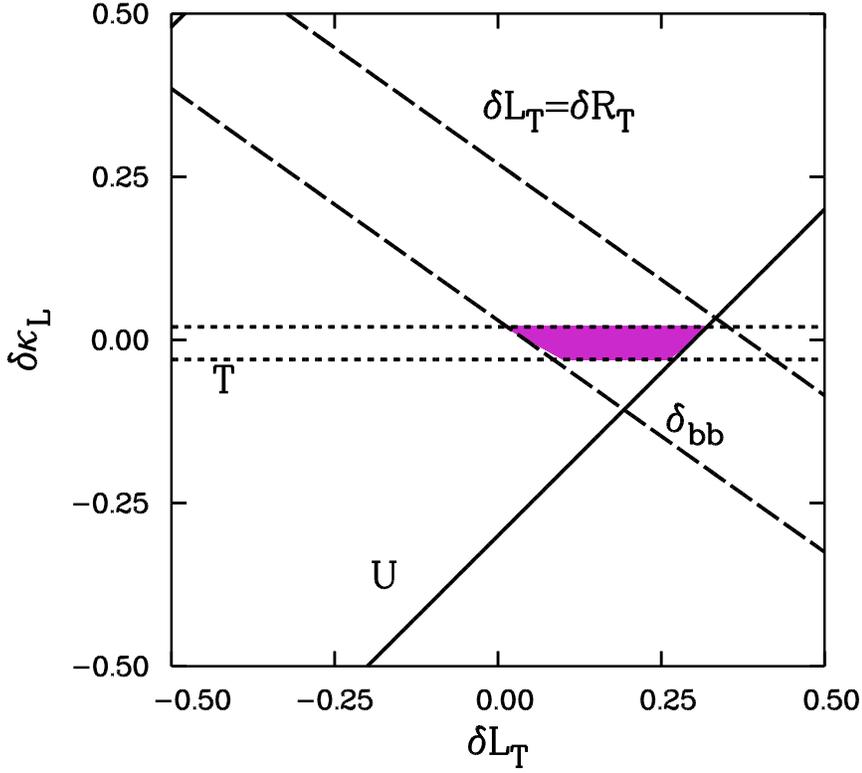}}}
\vspace{2.0in}
\caption[]{$90\%$ confidence level bounds from electroweak
data with $\mu=2M_t$.  The region enclosed by the  dotted and
dashed lines are the limits from  $T^{exp}_{\rm NEW}$ and
$\delta_{
b{\overline b},{\rm NEW}}^{exp}
$ respectively.  For $\delta L_t=\delta R_t$,
 the limit from $S^{exp}_{\rm NEW}$ is
$-2< \delta L_t <1$.  The region above the solid  line
is allowed by $U^{exp}_{\rm NEW}$. The region allowed by all
measurements is shaded.}
\end{figure}

\begin{figure}
\hspace{2.in}{\epsffile{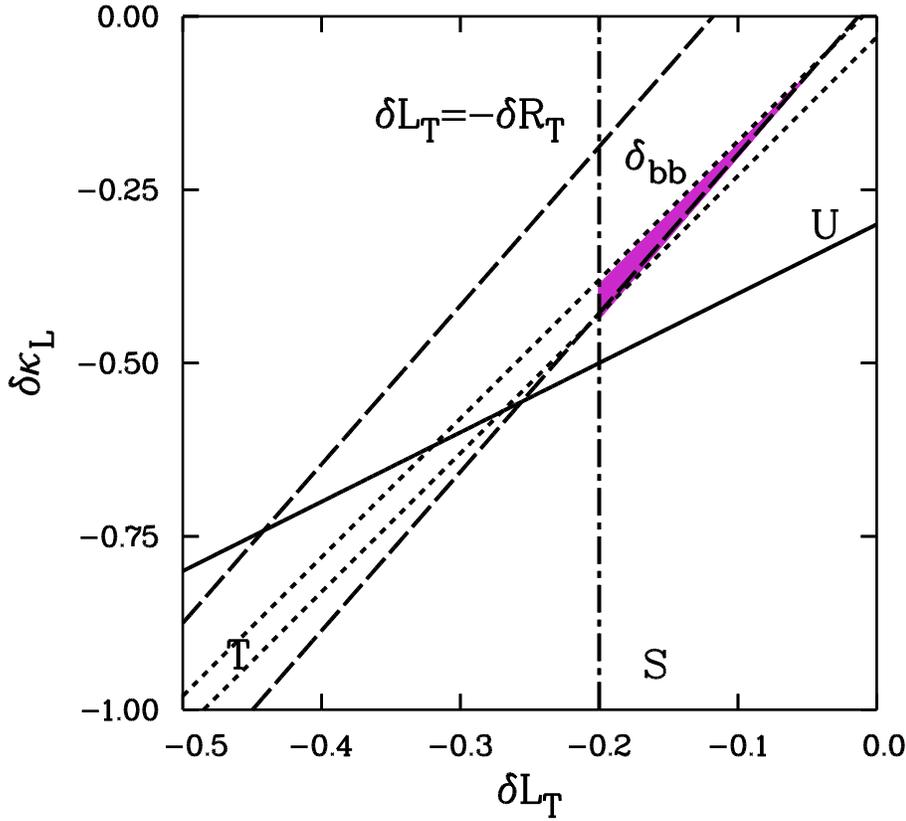}}
\vspace{2.0in}
\caption[]{$90\%$ confidence level bounds from electroweak
data with $\mu=2 M_t$.  The regions enclosed by the  dotted and
dashed lines are the limits from  $T^{exp}_{\rm NEW}$
 and $\delta_{
b{\overline b},{\rm NEW}}^{exp}$ respectively.
The region to the right of the dot-dashed line is
allowed from $S^{exp}_{\rm NEW}$, while the region above the
solid line is allowed by $U^{exp}_{\rm NEW}$.
}
\end{figure}

\begin{figure}
\centerline{\epsffile{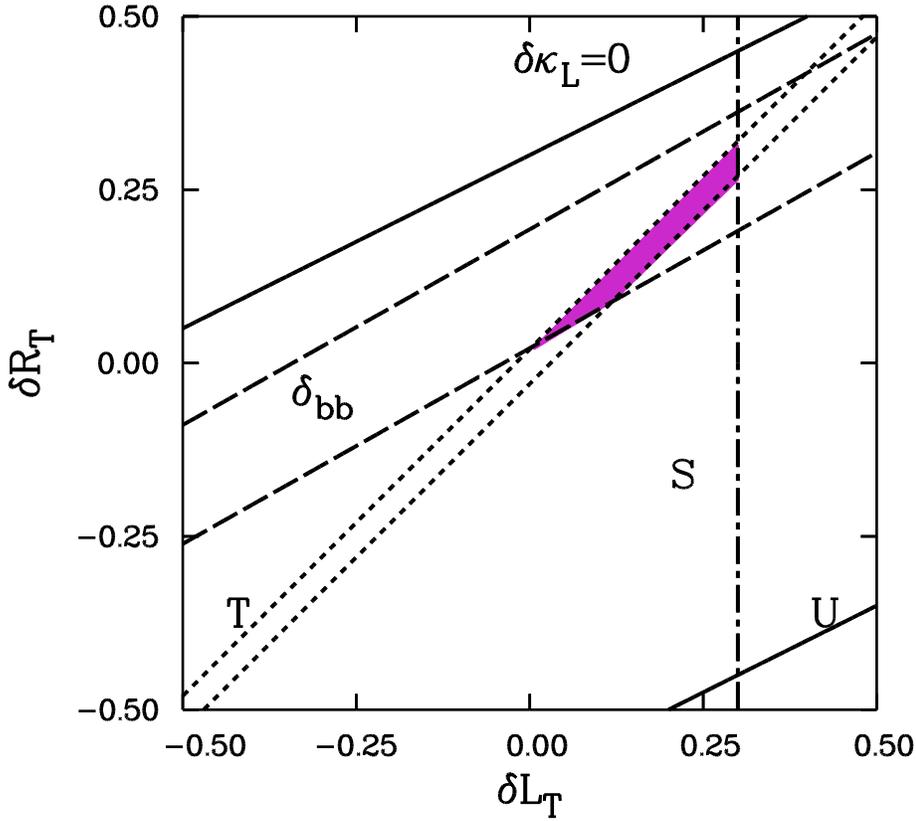}}
\vspace{2.0in}
\caption[]{$90\%$ confidence level bounds from electroweak
data with $\mu=2 M_t$.
The regions enclosed by the   dotted and
dashed lines are the limits from
 $T^{exp}_{\rm NEW}$
 and $\delta_{
b{\overline b},{\rm NEW}}^{exp}$ respectively.
The region to the left of the dot-dashed line is
allowed from $U^{exp}_{\rm NEW}$, while the region
below the solid line is that allowed by
$S^{exp}_{\rm NEW}$.  The region allowed by all
measurements is shaded.}
\end{figure}

In Fig. 1, we take $\delta L_t=\delta R_t$ and show the limits
on $\delta \kappa_L$ as a function of $\delta L_t$.
Including the limit from $U^{exp}_{\rm NEW}$
excludes a considerably larger
region of parameter space than would be excluded from
$T^{exp}_{\rm NEW}$ and $\delta_{b{\overline b},
{\rm NEW}}
^{exp}$ alone.
If the new couplings modify only the axial coupling to the $Z$,
${\cal L}
\sim \overline{t}\gamma^\mu \gamma_5 t Z_\mu$, then $\delta L_t=-
\delta R_t$.  We show the limits in this case in Fig. 2
and see that only a small region of parameter space is allowed.
 This
agrees with previous observations that the electroweak data tend
to prefer $\delta L_t=\delta R_t$ \cite{yuan}.
In Fig. 3, we set $\delta \kappa_L=0$ and show the $90\%$\
confidence level bounds on $\delta L_t$ and $\delta R_t$.
In this figure, we again see the
trend that in the allowed region, $\delta L_t\sim\delta R_t$.
In all three  figures, it is interesting to note the role of the
limit on $U^{exp}_{\rm NEW}$, which is more significant than
the limit from $S^{exp}_{\rm NEW}$.

  The effects of a possible right-handed coupling of the $W$ to
the $b$ and $t$ first arises at ${\cal O}(\delta \kappa_R^2)$
in $Z$ decays.
The best limit comes from $T_{\rm NEW}^{exp}$ and is
\beq
\mid \kappa_R \mid < .3
\quad .  \eeq
This is considerably weaker than the limit from the CLEO
measurement of $b\rightarrow s\gamma$\cite{rare}.
\beq
-.05 < {\delta \kappa_R\over \mid V_{tb}\mid} < .01 \quad .
\eeq
\section{Conclusions}

The increasing precision of electroweak measurements, along with the
recent measurement of the top quark mass, allows limits
to be placed on the non-standard model couplings of the top
quark to gauge bosons.
We have updated previous limits by including terms not enhanced by
the top quark mass.  These additional terms
allow a larger region of parameter space to be
excluded than in previous studies.

 It is interesting to speculate  as to the size
of these coefficients in various models.  If the non-standard couplings
arise from loop effects then they might be expected to have a
size $\sim {\alpha\over \pi}\sim .002$, several orders of
magnitude smaller than the current limits.  In models where the non-standard
couplings arise from four-fermion operators at a high scale, the
coefficients have a typical size \cite{zhang},
\beq
\delta L_t\sim {3 g_A\over 8 \pi^2}{M_t^2\over \Lambda^2}
\log\biggl({\Lambda^2\over M_t^2}\biggr)  \quad .
\eeq
In the top color model of Ref. \cite{zhang}, $g_A\sim 4 \pi (.11)$ and so
a scale $\Lambda\sim 2~TeV$ would yield a coefficient, $\delta L_t
\sim 10^{-3}$.  Unfortunately,
unlike the case with the b-quark couplings,
 the limits on top quark couplings
do not yet seriously
 constrain model building.

\section*{Acknowledgements}

The work of G. V. was supported in part
by a DOE OJI award under contract number DE-FG02-92ER40730.


\begin{thebibliography}{999}

\bibitem{ttbar}{W.~Bardeen, C.~Hill, and M.~Lindner,
{\it Phy. Rev.} {\bf D41} (1990) 1647; C. ~Hill, {\it Phys.
Lett.} {\bf B266} (1991) 419; S.~Martin, {\it Phys. Rev.}
{\bf D46} (1992) 2197; {\it Phys. Rev.} {\bf D45}(1992) 4283;
{\it Nucl. Phys.} {\bf B398} (1993) 359;
M.~Lindner and D.~Ross, {\it Nucl. Phys.} {\bf B370} (1992) 30;
R.~Bonisch, {\it Phys. Lett.} {\bf B268} (1991) 394;
C.~Hill, D.~Kennedy, T. Onogi, and H.~Yu, {\it Phys.
Rev.} {\bf D47} (1993) 2940}

\bibitem{hill}{C.~Hill, {\it Phys. Lett.} {\bf B345}
(1995) 483; E. ~Eichten and K.~Lane, BUHEP-95-11, hep-ph/9503433,
March, 1995.}

\bibitem{rare}{K.~Fukikawa and A.~Yamada, {\it Phys. Rev.}
{\bf D49} (1994) 5890; J.~Hewett and T.~Rizzo, {\it Phys.
Rev.} {\bf D49} (1994) 319.}

\bibitem{yuan}{K.~Whisnant, B.~L.~Young and X.~Zhang, {\it Phys. Rev.}
{\bf D52} (1995) 3115;
E.~Malkawi and C.-P. ~Yuan, {\it Phys. Rev.}
{\bf D50} (1994) 4462; D.~Carlson, E.~Malkawi, and C.-P. ~Yuan,
{\it Phys. Lett.} {\bf B337} (1994) 145.}

\bibitem{shif}{M.~Shifman, {\it Mod. Phys. Lett.} {\bf A10} (1995) 605.}

\bibitem{as}{S.~ Catani, DFF- 211/10/94, hep-ph/9411361, November, 1994;
P. Chankowski and S.~Pokorski, IFT-UW-95/5, hep-ph/9505304,
May, 1995;
 G.~Kane, R.~Stuart, and J.~Wells, UM-TH-95-16,
 hep-ph/9505207, April, 1995.}

\bibitem{lang}{J.~Erler and P.~Langacker, {\it Phys. Rev.}
{\bf D50} (1994) 1304.}


\bibitem{pz}{R.~Peccei and X.~Zhang, {\it Nucl. Phys.} {\bf B337}
(1990) 269; R.~Peccei, S.~Peris, and X.~Zhang, {\it Nucl.
Phys.} {\bf B349} (1991) 305.}

\bibitem{dv}{S.~Dawson and G.~Valencia, {\it Nucl. Phys.}
{\bf B439} (1995) 3.}

\bibitem{georgi}{A.~Manohar and H.~Georgi, {\it Nucl.
Phys. } {\bf B234} (1984) 189.}

\bibitem{zbb}{R. Holdom, R.~Chivukula, S.~Selipsky, and E.~Simmons,
{\it Phys. Rev. Lett.} {\bf 69} (1992) 575;
 R.~Chivukula,  E.~Simmons, and J.~Terning,
{\it Phys. Lett.} {\bf B331} (1994) 383;
G.-H. Wu,{\it Phys. Rev. Lett.} {\bf 74}
(1995) 4137; D.~Schaile and P.~Zerwas, {\it Phys. Rev.} {\bf 45}
(1992) 3263.}

\bibitem{longo}{A. Longhitano, {\it Nucl. Phys.} {\bf B188} (1981) 118.}

\bibitem{zhang}{ X. Zhang,
{\it Mod. Phys. Lett.} {\bf A9} (1994) 1955;
 C.~Hill and X. Zhang,
{\it Phys. Rev. } {\bf D51} (1995) 3563.}

\bibitem{stu}{D.~C.~Kennedy and B.~W.~Lynn, {\it Nucl. Phys.}
{\bf 322} (1989) 1;
M.~Peskin and T.~Takeuchi,
{\it Phys. Rev. Lett.} {\bf 65} (1990) 964;
{\it Phys. Rev.} {\bf D46} (1992) 381;
G.~Altarelli, and R.~ Barbieri, and F.~ Caravaglios,
{\it Nucl. Phys.} {\bf 405} (1993) 3.}

\end{thebibliography}
\end{document}